\documentstyle[12pt]{article}

\catcode`\@=11
\long\def\@makefntext#1{
\protect\noindent \hbox to 3.2pt {\hskip-.9pt
$^{{\ninerm\@thefnmark}}$\hfil}#1\hfill}	
\def\@makefnmark{\hbox to 0pt{$^{\@thefnmark}$\hss}} %ORIGINAL

\def\ps@myheadings{\let\@mkboth\@gobbletwo \def\@oddhead{\hbox{}
\rightmark\hfil\ninerm\thepage}
\def\@oddfoot{}\def\@evenhead{\ninerm\thepage\hfil
\leftmark\hbox{}}\def\@evenfoot{}
\def\sectionmark##1{}\def\subsectionmark##1{}}

\renewcommand{\thefootnote}{\fnsymbol{footnote}}

\newcounter{sectionc}\newcounter{subsectionc}\newcounter{subsubsectionc}
\renewcommand{\section}[1] {\vspace*{0.6cm}\addtocounter{sectionc}{1}
\setcounter{subsectionc}{0}\setcounter{subsubsectionc}{0}\noindent
{\normalsize\bf\thesectionc. #1}\par\vspace*{0.4cm}}
\renewcommand{\subsection}[1] {\vspace*{0.6cm}
\addtocounter{subsectionc}{1}
\setcounter{subsubsectionc}{0}\noindent
{\normalsize\it\thesectionc.\thesubsectionc. #1}\par\vspace*{0.4cm}}
\renewcommand{\subsubsection}[1]
{\vspace*{0.6cm}\addtocounter{subsubsectionc}{1}
\noindent {\normalsize\rm\thesectionc.\thesubsectionc.\thesubsubsectionc.
#1}\par\vspace*{0.4cm}}

\newcounter{appendixc}
\newcounter{subappendixc}[appendixc]
\newcounter{subsubappendixc}[subappendixc]

\renewcommand{\appendix}[1] {\vspace*{0.6cm}
\refstepcounter{appendixc}
\setcounter{figure}{0}
\setcounter{table}{0}
\setcounter{equation}{0}
\renewcommand{\thefigure}{\Alph{appendixc}.\arabic{figure}}
\renewcommand{\thetable}{\Alph{appendixc}.\arabic{table}}
\renewcommand{\theappendixc}{\Alph{appendixc}}
\renewcommand{\theequation}{\Alph{appendixc}.\arabic{equation}}
\noindent{\bf Appendix \theappendixc #1}\par\vspace*{0.4cm}}

\def\abstracts#1{{
\centering{\begin{minipage}{12.2truecm}\footnotesize\baselineskip=12pt\noindent
\centerline{\footnotesize ABSTRACT}\vspace*{0.3cm} \parindent=0pt #1
\end{minipage}}\par}}

\renewenvironment{thebibliography}[1]
{\begin{list}{\arabic{enumi}.}
{\usecounter{enumi}\setlength{\parsep}{0pt} 
\setlength{\leftmargin 1.25cm}{\rightmargin 0pt} 
\settowidth {\labelwidth}{#1.}\sloppy}}{\end{list}}

\newcounter{itemlistc}
\newcounter{romanlistc}
\newcounter{alphlistc}
\newcounter{arabiclistc}

\newcommand{\fcaption}[1]{
\refstepcounter{figure}
\setbox\@tempboxa = \hbox{\footnotesize Fig.~\thefigure. #1} \ifdim
\wd\@tempboxa > 6in
{\begin{center}
\parbox{6in}{\footnotesize\baselineskip=12pt Fig.~\thefigure. #1}
\end{center}}
\else
{\begin{center}
{\footnotesize Fig.~\thefigure. #1}
\end{center}}
\fi}

\newcommand{\tcaption}[1]{
\refstepcounter{table}
\setbox\@tempboxa = \hbox{\footnotesize Table~\thetable. #1} \ifdim
\wd\@tempboxa > 6in
{\begin{center}
\parbox{6in}{\footnotesize\baselineskip=12pt Table~\thetable. #1}
\end{center}}
\else
{\begin{center}
{\footnotesize Table~\thetable. #1}
\end{center}}
\fi}

\def\@citex[#1]#2{\if@filesw\immediate\write\@auxout
{\string\citation{#2}}\fi
\def\@citea{}\@cite{\@for\@citeb:=#2\do
{\@citea\def\@citea{,}\@ifundefined
{b@\@citeb}{{\bf ?}\@warning
{Citation `\@citeb' on page \thepage \space undefined}} {\csname
b@\@citeb\endcsname}}}{#1}}

\newif\if@cghi
\def\cite{\@cghitrue\@ifnextchar [{\@tempswatrue
\@citex}{\@tempswafalse\@citex[]}}
\def\citelow{\@cghifalse\@ifnextchar [{\@tempswatrue
\@citex}{\@tempswafalse\@citex[]}}
\def\@cite#1#2{{$\null^{#1}$\if@tempswa\typeout
{IJCGA warning: optional citation argument ignored: `#2'} \fi}}

 1
 1
 1

\font\ninerm=cmr9

\def\ie{{\em i.e.}}
\def\be{\begin{equation}}
\def\ee{\end{equation}}
\def\bea{\begin{eqnarray}}
\def\eea{\end{eqnarray}}
\def\ie{{\em i.e.}}

\newcommand{\RR}{{\mbox{{\bf R}}}}

\newcommand{\topo}{{\scriptstyle {\rm top}}}

\begin{document}

\begin{flushright}
US-FT-3-97 \\
hep-th/9704032 
\end{flushright}
\vskip0.5cm

\centerline{\normalsize\bf DUALITY AND TOPOLOGICAL QUANTUM FIELD
THEORY\footnotemark} \baselineskip=22pt
%\centerline{\normalsize\bf GUIDELINES FOR TYPESETTING A CAMERA-READY}
%\baselineskip=16pt
%\centerline{\normalsize\bf MANUSCRIPT BY COMPUTER}

\footnotetext{Talk given at the ``Workshop on Frontiers of Field Theory,
Quantum Gravity and String Theory"  held in Puri, India,  on December 12-21,
1996}

%\vfill
%\vspace*{0.6cm}
\centerline{\footnotesize J. M. F. Labastida and M. Mari\~no} 
\baselineskip=13pt
\centerline{\footnotesize\it Departamento de F\'\i sica de Part\'\i culas}
\baselineskip=12pt
\centerline{\footnotesize\it Universidade de Santiago de Compostela}
\baselineskip=12pt
\centerline{\footnotesize\it E-15706 Santiago de Compostela, Spain}
\centerline{\footnotesize E-mail: labastida@gaes.usc.es}
\centerline{\footnotesize E-mail: marinho@gaes.usc.es} \vspace*{0.3cm}
%\centerline{\footnotesize and}
%\vspace*{0.3cm}
%\centerline{\footnotesize SECOND AUTHOR'S NAME} %\baselineskip=13pt
%\centerline{\footnotesize\it Group, Company, Address, City, State ZIP/Zone,
%Country}

%\vfill
\vspace*{0.3cm}
\abstracts{We present a summary of the applications of duality to
Donaldson-Witten theory and its generalizations. Special emphasis is made on
the computation of Donaldson invariants in terms of Seiberg-Witten invariants
using recent results in $N=2$ supersymmetric gauge theory. A brief account on
the invariants obtained in the theory of non-abelian monopoles is also
presented.}

\vspace*{0.6cm}
\normalsize\baselineskip=15pt
\setcounter{footnote}{0}
\renewcommand{\thefootnote}{\alph{footnote}}

%\section{Introduction}

Topological quantum field theories\cite{tqft,thompson,moore} (TQFTs) have provided
very powerful tools to address problems related to the topology of low-dimensional
manifolds.  They involve field theory representations of some of the most recent
topological invariants studied on these manifolds. In three dimensions
Chern-Simons gauge theory\cite{csgt} has led to a description of polynomial
invariants for knots and links which generalize the Jones polynomial\cite{jones}.
In four dimensions, the TQFT known as Donaldson-Witten
theory\cite{tqft} has given a new way to look at Donaldson
invariants\cite{donald}. The connection between these theories and these two types
of invariants was achieved studying the corresponding quantum field theories in
their non-perturbative and perturbative regimes, respectively. It was clear from
the first works in the field that an analysis of these theories in other regimes
would certainly provide new points of view on these invariants.

Chern-Simons gauge theory was studied from a perturbative point
of view\cite{gmm,natan} soon after its formulation by Witten in 1988. The
perturbative series  provided an infinite sequence of numerical knot invariants
which  were later identified with Vassiliev
invariants\cite{bilin,kont,barnatan}. Vassiliev invariants were formulated from
a mathematical point of view\cite{vass} basically at the same time. It took,
however, some time until a relation between them was pointed out. Nevertheless,
Chern-Simons gauge theory has provided an integral representation of at least a
subset of Vassiliev invariants\cite{alla}. Furthermore, the richness of quantum
field theory is being exploited to extend these invariants for the case of
links\cite{esther} and most likely, through the analysis of the theory in
different gauges, it will provide alternative representations.

In four dimensions, the analysis of Donaldson-Witten theory in a regime
different than the perturbative one seemed a rather unsurmountable 
problem before
1994. However, in that year Seiberg and Witten unraveled the strong-coupling
behavior of $N=2$ supersymmetric gauge theory using arguments based on
duality\cite{sw}. Their results could certainly be applied to Donaldson-Witten
theory since this theory can be regarded as a twisted form of $N=2$
supersymmetric gauge theory. Indeed this  was done by Witten\cite{abm} in the fall
of 1994 obtaining an unexpected new representation for Donaldson invariants. In
this representation Donaldson invariants are expressed in terms of a new type of
invariants now known as Seiberg-Witten invariants. These invariants turn out to
be simpler  than Donaldson invariants and, presumably, more powerful. They have
certainly opened a new point of view on the topology of four-manifolds.

In this paper we will concentrate on Donaldson-Witten theory. The aim is
to account for the progress made after the appearance of Seiberg-Witten
solution for $N=2$ supersymmetric gauge theories. Before entering into details
let us briefly  describe how ideas based on duality, intrinsic in this
solution, can be used in the contest of Donaldson-Witten theory.

Donaldson-Witten theory can be formulated in terms of a twisted version of $N=2$
supersymmetric gauge theory or, from a more geometrical point of view, in terms
of a representative of the Thom class of a vector bundle associated to certain
moduli problem in the framework of the Mathai-Quillen formalism\cite{jeffrey}.
Twisted $N=2$ supersymmetric theories, in general, are associated to certain
moduli problems which, properly treated in the contest of the Mathai-Quillen
formalism, lead to representatives of the Thom class which are the exponential of
the twisted action. Both pictures of Donaldson-Witten theory have been known for
some time. One important property of the resulting TQFT is that the vevs. of 
its observables are independent of the coupling constant. This
means that these vevs. could be computed in either the strong or the weak
coupling limit. The weak coupling limit analysis showed  the relation
of the observables of the theory to Donaldson invariants. However, in
such analysis no new progress was made from the quantum field theory
representation in what respect to the calculation of these invariants. The
difficult problems that one has to face are similar to the ones
in ordinary Donaldson theory. 

In 1994 Seiberg and Witten worked out the strong coupling limit behavior of
$N=2$ supersymmetric Yang-Mills theory\cite{sw}. One would expect that the twisted
version of the corresponding effective theory would be related to
Donaldson-Witten theory. Furthermore, since in the TQFT observables are independent of the coupling constant, the weak coupling
limit of the effective theory should be exact, \ie, it would lead to
Donaldson invariants. This is in fact what turns out to be the case. The twisted
effective theory could be regarded as a TQFT {\it
dual} to the original one. In addition, one could ask for the {\it dual} moduli
problem associated to this dual TQFT. It turns out that
this moduli space is an abelian version of the moduli space of instantons
modified by the presence of chiral spinors. This space is known as the moduli
space of abelian monopoles\cite{abm}. Being related to an abelian gauge theory
this space is simpler to analyze than the moduli space of instantons.
Furthermore,  for a large set of four-manifolds
(of simple type) only particular classes of abelian gauge fields (basic classes)
contribute. For these classes the moduli space of abelian monopoles reduces to a
finite set of points. 

Generalizations of Donaldson-Witten theory introducing matter fields have been
studied for some time\cite{rocek,top,ans}. Recently, the resulting theories have
been understood in the Mathai-Quillen formalism\cite{abmono,nabm} and their
non-perturbative analysis has been carried out, at least for the gauge group
$SU(2)$ and one multiplet of matter in the fundamental representation. From a
perturbative point of view the associated moduli space corresponds to
non-abelian monopoles. The set of observables of this theory has  a  structure
which is similar  to the one in standard Donaldson-Witten
theory and, although they are based in a different moduli space, it turns out
that they can also be expressed in terms of Seiberg-Witten invariants. There
seems to exist a set of types of topological invariants which can be expressed
\cite{last} in terms of these invariants. 

After a brief introduction to TQFT in the first section we will describe
in the second section 
how Donaldson invariants are computed in Donaldson-Witten theory
using recent results in $N=2$ supersymmetric gauge theory. In section
three we present a brief description of the theory of non-abelian monopoles
as well as some concluding remarks.

\section{Topological Quantum Field Theory}
To be specific, we will restrict ourselves to a particular set of TQFTs, namely those with a gauge symmetry and a BRST operator 
$Q$. These theories are then of cohomological type and are usually constructed 
starting from a supersymmetric gauge theory. Both Donaldson-Witten theory and 
the Seiberg-Witten theory of abelian monopoles are theories of this type, as 
well as the non-abelian monopole theory. The two key points in these kinds 
of theories are the mathematical interpretation associated to them and the 
relation to the underlying physical model. We will begin with a rather general  
exposition of the first aspect, and then we will discuss the twisting 
procedure which makes the connection with $N=2$ supersymmetric gauge theories. 
More details can be found in references\cite{thompson,moore,tesis}. 

Topological quantum gauge theories of the cohomological type are characterized by 
the following data. First of all we have a configuration space ${\cal X}$, 
whose elements are fields ${\phi_i}$ defined on some Riemannian manifold 
$M$. Among these 
fields there is the gauge connection associated to a principal $G$-bundle $P$ 
over $M$, $A$. The group of gauge transformations ${\cal G}$ acts on 
${\cal X}$ according to the different gauge-structure of the fields. For 
instance, on the gauge connections the gauge transformations are the standard 
ones, and if there are matter fields in ${\cal X}$, ${\cal G}$ acts on them 
according to the representation of the gauge group we choose. In this case, 
there is also a representation bundle $E$ on $M$ associated to $P$. Notice that 
in general the configuration space ${\cal X}$ is infinite-dimensional.
 
The next ingredient in the construction is an (infinite-dimensional) vector 
space ${\cal F}$ also with an action of the group of 
gauge transformations. This vector space corresponds, as we will see in a 
moment, to the quantum numbers of the field equations. One can then 
consider the configuration space ${\cal X}$ as a principal 
${\cal G}$-bundle and ${\cal F}$ as a representation space and construct the 
associated vector bundle, 
\be
{\cal E}= {\cal X} \times_{\cal G} {\cal F},
\label{fibrao}
\ee
which will be the fundamental geometrical object in this class of theories. 
The last ingredient is a ${\cal G}$-equivariant map,
\be
s: {\cal X} \rightarrow {\cal F}, \quad s(g\cdot \phi_i)=g \cdot s(\phi_i), 
\label{seccion}
\ee
which descends to a section ${\hat s}: {\cal X}/{\cal G} 
\rightarrow {\cal E}$. The zero locus of the section ${\hat s}$ is the {\it 
moduli space} ${\cal M}$ associated to the problem, and its elements 
are precisely the 
fields that verify the equation $s(\phi_i)=0$, modulo gauge transformations. 
This is already a familiar object in field theory. Instanton equations, the 
Bogomolny monopole equations or the Dirac equation can be all reinterpreted 
in this framework. Although we started from infinite dimensional spaces, 
the moduli space ${\cal M}$ will be finite-dimensional provided that the map  
$s$ verifies some conditions (ellipticity), and in this case the dimension 
of the moduli space can be generically obtained from an index theorem. 

The topological charge can be understood in this context as follows: as both 
${\cal X}$, ${\cal F}$ have a ${\cal G}$ action, there is a correspondence 
between elements in the Lie algebra of ${\cal G}$, $\hbox{\rm Lie}({\cal G})$,
and  vector fields on ${\cal X}$, ${\cal F}$, given by a map $C$. This map, 
geometrically, gives the fundamental vector fields on the principal ${\cal G}$-
bundle ${\cal X}$, and physically corresponds to the gauge slices in the 
space of fields. One can then construct the operator,
\be
Q=d-\phi^a \iota (C(T_a)),
\label{cu}
\ee
where $d$ is the de Rham operator, $T_a$ is a basis of the Lie algebra of 
$G$, and $\phi^a$ are the components of an arbitrary field $\phi \in 
\hbox{\rm Lie}({\cal G})$  w.r.t. the basis $T_a$ 
(not to be confused with the general fields 
introduced before). $\iota$ refers as usual to the 
inner product with a vector field. This operator gives precisely 
the ${\cal G}$-equivariant cohomology of the configuration space 
${\cal X}$, and 
the cohomology classes thus obtained are in one-to-one correspondence with 
the cohomology of ${\cal X}/{\cal G}$. These cohomology classes are called 
the {\it observables} of the TQFT. 

What can one compute in this context? The first quantity of topological 
interest arises when the dimension of the moduli space is zero. In the 
finite-dimensional case this happens (for generic, transversal sections)
when the rank of ${\cal E}$ equals the dimension of the base 
space ${\cal X}/{\cal G}$. In this case, one can compute the Euler 
characteristic of the vector bundle ${\cal E}$ by integrating its 
Euler class $\chi({\cal E})$ on the base manifold. The Euler class can 
be obtained as the pullback of the Thom class of ${\cal E}$, ${\Phi}({\cal E})$ 
using the section ${\hat s}$. One then has for the Euler characteristic the 
expression,
\be
Z=\int_{{\cal X}/{\cal G}}[d \phi_i] \,\ {\hat s}^*({\Phi}({\cal E})), 
\label{parti}
\ee
where the fields in the measure are defined modulo gauge transformations. 
If the dimension of the moduli space is non-zero, there are other 
interesting quantities, the intersection numbers in the moduli space 
${\cal M}$. Consider then a basis for the cohomology ring 
of ${\cal X}/{\cal G}$, given by the observables of the theory 
${\cal O}_a$. If we group all the observables in a generating function for the 
intersection numbers, we get:
\be
\langle \hbox{\rm exp}(\sum_{a} \alpha_a {\cal O}_a ) \rangle = 
\int_{{\cal M}} \hbox{\rm exp}(\sum_{a} \alpha_a i^{*}{\cal O}_a ),
\label{generador}
\ee
where $i : {\cal M} \hookrightarrow {\cal X}/{\cal G}$ is the inclusion of the 
moduli space into the configuration space modulo gauge transformations. 
But it is a standard result in algebraic topology that ${\chi}({\cal E})$ is the 
Poincar\'e dual of the zero locus ${\cal M}={\hat s}^{-1}(0)$. One then obtains 
\be
\langle \hbox{\rm exp}(\sum_{a} \alpha_a {\cal O}_a ) \rangle = 
\int_{{\cal X}/{\cal G}}[d \phi_i] \,\ {\chi}({\cal E}) \wedge 
\hbox{\rm exp}(\sum_{a} \alpha_a {\cal O}_a ).
\label{vevs}
\ee
From this expression one sees that, in the expansion of the exponential, there 
will be contributions only for those products of observables whose total 
degree (as differential forms) equals the dimension of ${\cal M}$. This is 
a selection rule for the correlation functions of the observables, and has 
a Field Theory description: the degree of the forms is the quantum number 
associated to an $U(1)_{\cal R}$ symmetry, and the selection rule is the 
`t Hooft condition.

To make the connection with Field Theory, we still need more ingredients. 
Although we won't give the details here, the crucial step involves the 
introduction of a particular representative for the Thom class due to 
Mathai and Quillen\cite{mathai}. The Mathai-Quillen formalism 
has been crucial to elucidate 
the geometric structure of Cohomological Gauge Theories, and its importance 
in this context was first realized by Atiyah and Jeffrey in
reference\cite{jeffrey}  and further clarified in reference\cite{moore}. Let us
see how to obtain a  field theory description of the topological invariants
introduced above.
 Denote by ${\psi_i}$ a set of fields associated to the fibre 
${\cal F}$ and to the Lie algebra of the group of gauge transformations 
$\hbox{\rm Lie}({\cal G})$. We also introduce Grassmannian fields associated 
to the integration of differential forms on the spaces involved. We will 
denote them by ${\hat \phi}_i$, ${\hat \psi}_i$. Physically they correspond to 
the superpartners of the fields ${ \phi}_i$, ${ \psi}_i$. The result of 
\cite{jeffrey,moore} 
can be summarized in the following equation:
\be
\pi ^* ({\chi}({\cal E})) \Phi({\cal X} \rightarrow {\cal X}/{\cal G})= 
\int [d \psi ]_i [d {\hat \psi}]_i \,\  
\hbox{\rm exp}(-S_\topo[\phi_i,\psi_i, {\hat \phi}_i, {\hat \psi}_i]).
\label{mq}
\ee
In this equation, $\pi$ is the projection map of the principal ${\cal G}$-
bundle ${\cal X}$, and $\Phi({\cal X} \rightarrow {\cal X}/{\cal G})$ is a 
projection form that allows one to lift the 
integration on ${\cal X}/{\cal G}$ to the full principal bundle ${\cal X}$. 
$S_\topo[\phi_i,\psi_i, {\hat \phi}_i, {\hat \psi}_i]$ is the action 
of a certain field theory, the TQFT associated 
to the moduli problem. Notice that 
the term in the l.h.s. is a differential form on the bundle ${\cal X}$, and 
depends on the fields $\phi_i$ and its superpartners ${\hat \phi}_i$. These 
superpartner fields are to be understood as a basis of 
differential forms for the 
``coordinate" fields $\phi_i$. Taking this into account, we can write the 
Euler characteristic and the intersection numbers as,
\bea
Z&=& \int_{\cal X}[d \phi_i][d {\hat \phi}_i][d \psi_i][d {\hat \psi}_i]
  \,\ \hbox{\rm exp}(-S_\topo[\phi_i,\psi_i, {\hat \phi}_i, {\hat \psi}_i]), 
\nonumber\\
\langle \hbox{\rm exp}(\sum_{a} \alpha_a {\cal O}_a ) \rangle &=& 
\int_{\cal X}[d \phi_i] [d {\hat \phi}_i][d \psi_i][d {\hat \psi}_i] \,\ 
\hbox{\rm exp}(-S_\topo[\phi_i,\psi_i, {\hat \phi}_i, {\hat \psi}_i]+
\sum_{a} \alpha_a {\cal O}_a).
\label{basica}
\eea
In this way, we already have the fundamental bridge between the moduli problem 
and the field theory framework: associated to a moduli problem there is a 
certain TQFT with action $S_\topo$. The Euler 
characteristic of the bundle involved in the moduli problem is the 
partition function of the TQFT, and the intersection 
numbers on the moduli space are the correlation functions of the observables 
${\cal O}_a$. There are two important properties about the topological action 
appearing here, and both can be derived in the framework of the Mathai-Quillen 
formalism. The first one is that the bosonic part of $S_\topo$ has the 
form,
\be 
S_\topo= ||s(\phi_i)||^2 + \cdots, 
\label{seccionbos}
\ee
where the norm is taken w.r.t. an appropriate ${\cal G}$-invariant 
metric on the vector space ${\cal F}$. The second property has been crucial to 
understand the properties of TQFTs. The $Q$ operator
can  be extended in a natural way to act on all the fields in the theory. On the 
configuration space ${\cal X}$ it is defined as in (\ref{cu}). The fields 
$\psi_i$, ${\hat \psi}_i$ are fields in the vector space ${\cal F}$ or in the 
Lie algebra $\hbox{\rm Lie}({\cal G})$. In the first case, $Q$ acts as in
(\ref{cu})  as well, and the action on the fields in the Lie algebra is 
just the coadjoint 
action. With the $Q$ operator extended in this way, one has a BRST operator for 
the TQFT defined by $S_\topo$. As ${\cal O}_a$ 
and $S_\topo$ are representatives of (equivariant) 
cohomology classes, one has:
\be
[Q, {\cal O}_a \}=[Q, S_\topo]=0.
\label{cerrados}
\ee
Furthermore, in almost all the TQFTs of the
 cohomological type, one also has 
\be
S_\topo= \{Q, V \},
\label{ex}
\ee
where $V$ is a certain functional of the fields usually called gauge fermion. 
The fact that $S_\topo$ is $Q$-exact does not mean that 
$S_\topo$ is trivial. Recall that, because of (\ref{mq}), the topological 
action is essentially a characteristic class, and these can be locally 
trivialized in terms of secondary characteristic classes. This also allows one 
to prove that characteristic classes are truly topological, and they don't 
depend on the metric or connection that one chooses to obtain a representative. 
Namely, using this local exactness 
one can easily prove that the 
variation of a characteristic classes w.r.t. a 
change in the connection is just an exact form. 
In the same way, if one considers the change of the topological action w.r.t. 
to a change in the metric of the underlying manifold $M$, $g_{\mu \nu}$ (which 
generically will explicitly appear in $S_\topo$), one obtains from 
(\ref{ex})
\be
T_{\mu \nu} = {\delta S_\topo \over \delta g^{\mu \nu}}= 
\{Q, {\delta V \over \delta g^{\mu \nu}}\}.
\label{energia}
\ee
The energy-momentum tensor is then a BRST commutator. 

After this review of the mathematical framework of topological quantum gauge
theories  of the cohomological type, we will briefly address the most useful
procedure  to obtain them from $N=2$ supersymmetric gauge theories: the twist. 
In this procedure the starting point is an $N=2$ supersymmetric quantum
field theory. The basic ingredient consists of extracting a 
scalar
symmetry out of the $N=2$ supersymmetry. Let us consider the case of $d=4$. In
${\RR}^4$ the global symmetry group when $N=2$ supersymmetry is present is 
$H= SU(2)_L\otimes SU(2)_R \otimes SU(2)_I \otimes U(1)_{\cal R} $ where
${\cal K} = SU(2)_L \otimes SU(2)_R$ is the rotation group and 
$SU(2)_I \otimes U(1)_{\cal R}$ is the internal symmetry group. The
supercharges $Q^i_\alpha$ and $\overline Q_{i\dot\alpha }$ which generate
$N=2$ supersymmetry have the following transformations under $H$:
\begin{equation}
Q^i_\alpha \;\; (\frac{1}{2},0,\frac{1}{2})^1,
\;\;\;\;\;\;\;\;\;\;\;
\overline Q_{i \dot\alpha } \;\;
 (0,\frac{1}{2},\frac{1}{2})^{-1},
\label{loli}
\end{equation}
where the superindex denotes the $U(1)_{\cal R}$ charge and the numbers
within parentheses the representations under each of the factors in 
$SU(2)_L\otimes SU(2)_R \otimes SU(2)_I$.

The twist consists of considering as the rotation group the group
${\cal K}' = SU(2)_L'\otimes SU(2)_R$ where $SU(2)_L'$ is the diagonal subgroup
of $SU(2)_L\otimes SU(2)_I$. This implies that the isospin index $i$ becomes
a spinorial index $\alpha$: $Q^i_\alpha \rightarrow Q^\beta_\alpha$
and $\overline Q_{i\dot\beta } \rightarrow G_{\alpha\dot\beta}$. Precisely the
trace of $Q^\beta_\alpha$ is chosen as the generator of the scalar symmetry:
$Q = Q^\alpha_\alpha$. Under the new global group $H'={\cal K}'\otimes
U(1)_{\cal R}$, the symmetry generators transform as:
\begin{equation}
G_{\alpha\dot\beta} \;\; (\frac{1}{2},\frac{1}{2})^{-1},
\;\;\;\;\;\;\;\;\;\;
Q_{(\alpha\beta)} \;\; (1,0)^1,
\;\;\;\;\;\;\;\;\;\;
Q \;\; (0,0)^1.
\label{olga}
\end{equation}

Once the scalar symmetry is found we must study if, as stated in 
(\ref{energia}), the energy-momentum tensor is exact,  \ie, if it can be written
as the transformation of some quantity under $Q$. The $N=2$ supersymmetry
algebra gives a  necessary condition for this to hold. Notice that, after
the twisting, such an algebra becomes:
\begin{equation}
\{Q^i_\alpha, \overline Q_{j\dot\beta} \} = \delta^i_j P_{\alpha\dot\beta}
\longrightarrow
\{ Q , G_{\alpha\beta} \} = P_{\alpha\dot\beta},
\label{conchi}
\end{equation}
where $P_{\alpha\dot\beta}$ is the momentum operator of the theory.
Certainly (\ref{conchi}) is only a necessary condition for the theory being
topological. However, up to date, for all the $N=2$ supersymmetric models whose
twisting has been studied, the relation on the right hand side of
(\ref{conchi}) has become valid for the whole energy-momentum tensor.

\section{Donaldson-Witten Theory}

Donaldson-Witten theory was historically the first TQFT, 
and was constructed by Witten in reference\cite{tqft}. The geometrical 
framework for this model is the following. We consider a 
Riemannian four-manifold $X$ together with a principal $G$-bundle $P$. The 
configuration space is just the space of $G$-connections on this bundle, ${\cal 
A}$. The vector space ${\cal F}$ is the space of self-dual forms with 
values in the adjoint bundle ${\bf g}_P$, $\Omega^{2,+}({\bf g}_P)$. The 
map $s$ is given by
\be
s(A)=F^{+}_A, 
\label{asd}
\ee
where $F^{+}_A$ is the self-dual part of the curvature of the connection $A$. 
The moduli space associated to $s$ is then the moduli space of anti-self-dual 
(ASD) connections on $X$, {\it i.e.}, the moduli space of ASD $G$-instantons. 
This moduli space was considered by Atiyah, Hitchin and Singer in
reference\cite{ahs}  and is the building block of Donaldson theory\cite{donald}.
Indeed, the  intersection numbers defined in (\ref{generador}) are in this case
the  Donaldson invariants. We will be mainly concerned with the simplest 
case $G=SU(2)$. The moduli space of ASD connections has then the dimension 
\be 
\hbox{\rm dim} \,\ {\cal M}_{\rm ASD}=8k-{3 \over 2} (\chi + \sigma), 
\label{dim}
\ee
where $k$ is the instanton number, and $\chi$ and $\sigma$ are respectively 
the Euler characteristic and the signature of the four-manifold $X$. Although 
we haven't discussed the subtleties concerning the structure of moduli spaces, 
there is a crucial point which will be important in what follows. In order 
to avoid singularities in the moduli space, one has to be careful with abelian 
instantons that can be a solution to $F^{+}_A=0$ by embedding $U(1)$ 
in $SU(2)$. These abelian instantons are called the {\it reducible solutions}
 to the 
ASD equations. One can prove that, if the manifold $X$ is such that 
$b_2^{+} >1$, there are no reducible solutions. We will assume that this is 
the case in what follows. 
 
The cohomology associated to the operator $Q$ is in this 
case,
\be
\delta A_\mu = \psi_\mu,  \mbox{\hskip2cm}
\delta \psi = d_A\phi, \mbox{\hskip2cm}
\delta \phi = 0,
\label{titin}
\ee
where the Grassmannian field $\psi$ is understood as a basis of differential 
forms for the configuration space ${\cal A}$, and $\phi$ is a the generator of 
the ${\cal G}$-equivariant cohomology appearing in (\ref{cu}). The other fields 
in the theory include a Grassmannian self-dual $2$-form $\chi_{\mu \nu}$, with 
values in ${\bf g}_P$, which is associated 
to the fibre ${\cal F}$. One also has Lie algebra valued fields $\lambda$, 
$\eta \in \Omega^{0}(X, {\bf g}_P)$. To construct the cohomology 
ring of this model one uses the 
{\it descent procedure} introduced in\cite{tqft}. The starting point is 
the operator ${\cal O}^{(0)}={\cal O}$ given by
\be
{\cal O}={1 \over 8 \pi^2} \hbox{\rm Tr}(\phi^2), 
\label{four}
\ee
which is a differential form of degree four on the moduli space. 
One recursively finds $k$-forms on $X$, ${\cal O}^{(k)}$, which are 
$4-k$ forms on the moduli space, such that 
the following descent equations are verified:
\be
d{\cal O}^{(k)}= \{Q, {\cal O}^{(k+1)} \}.
\label{descent}
\ee
One then finds,
\be
{\cal O}^{(1)}={1 \over 4 \pi^2} \hbox{\rm Tr}(\phi\psi),
\mbox{\hskip1cm}
{\cal O}^{(2)}={1 \over 4 \pi^2} \hbox{\rm Tr}(\phi F+{1\over 2}\psi\wedge\psi),
\mbox{\hskip1cm}
{\cal O}^{(3)}={1 \over 4 \pi^2} \hbox{\rm Tr}(\psi \wedge F).
\ee
From the descent equations (\ref{descent}) follows that if $\Sigma$ is a 
$k$-dimensional homology cycle, then,
\be
I(\Sigma)= \int_{\Sigma} {\cal O}^{(k)},
\label{cerrao}
\ee
is in the cohomology of $Q$. For simply connected four-manifolds (the main 
focus of Donaldson theory), $k$-dimensional homology cycles only exist for 
$k=0,2,4$, and in the last case the observable $I(X)$ is just the instanton 
number. The generators of the cohomology ring of ${\cal A}/{\cal G}$ are then 
the observable ${\cal O}$ and the family of observables $I(\Sigma_a)$, where 
the $\Sigma_a$ denote a basis of the two-dimensional homology of $X$, and 
therefore $a=1, \cdots, \hbox{\rm dim} \,\ H_2(X, {\bf Z})$. The generating 
function corresponding to (\ref{generador}) will be in this case
\be
\langle \hbox{\rm exp}(\sum_{a} \alpha_a I(\Sigma_a) + \mu {\cal O}) \rangle,
\label{gendon}
\ee
and a sum over the instanton number is understood (as usual in Field Theory, 
we sum in the correlation functions over all the possible topological 
sectors). Notice that, as the differential forms on the moduli space 
appearing in the generating function 
are of even degree, (\ref{dim}) must be even to have non-vanishing 
correlation functions.

The TQFT associated to this moduli problem can be 
constructed with the Mathai-Quillen formalism or by a twisting of $N=2$ 
supersymmetric Yang-Mills theory. This is the point of view which will be 
useful to make contact with the Seiberg-Witten solution and then with 
the abelian monopole equations. The twisting procedure is standard, and one 
obtains (on-shell) the same fields we introduced before in the framework of 
the Mathai-Quillen formalism, as well as the same operator $Q$ (the topological 
charge). The field content of $N=2$ supersymmetric Yang-Mills theory is given 
by an $N=2$ vector multiplet with the following components: a 
one-form connection $A_{\mu}$, two Majorana spinors $\lambda_{i \alpha}$,
 $i=1,2$, a complex scalar $B$ and an auxiliary field $D_{ij}$ (symmetric in 
$i$ and $j$). The scalar field has an ${\cal R}$-charge $-2$, while 
the gluinos have ${\cal R}$-charge $-1$. After the twist, the fields 
become precisely the ones  in the Mathai-Quillen formulation.
 Notice that, as the twist changes the spin of the fields, all the fields 
in the twisted model are 
differential forms, hence they 
are well defined on any curved four-manifold $X$ (in contrast, 
the untwisted $N=2$ Yang-Mills theory 
has spinor fields, and then it is not well-defined on manifolds which are not 
${\rm Spin}$). The twisted action  coincides with the one obtained in the
framework of the Mathai-Quillen  formalism when one constructs a representative
for the Thom form. This  equivalence between the two formulations was first
discovered in  reference\cite{jeffrey}.

As we saw in the preceding section, Donaldson invariants are described by 
correlation functions of the twisted $N=2$ supersymmetric Yang-Mills theory, 
and because of (\ref{energia}) they are independent of the metric $g$ we choose 
on the four-manifold $X$. One can think then in evaluating the path-integral 
in the two limits $g \rightarrow 0$ (short distances, UV regime) and 
$g \rightarrow \infty$ (long distances, IR regime). As $N=2$ supersymmetric 
Yang-Mills theory is asymptotically free, it is weakly coupled in the UV. 
Indeed, the semiclassical approximation is exact because of the metric 
independence. In this approximation the path integral is localized on the 
moduli space of ASD instantons, $s(A)=0$, and we recover the usual description 
of Donaldson invariants in terms of intersection numbers in the moduli space. 
On the other hand, if one goes to the IR limit it is necessary to deal with 
a strongly-coupled gauge theory. Because of this, little progress 
has been made in understanding 
Donaldson theory from the physical point of view 
until Seiberg and Witten found the 
exact low-energy description of $N=2$ 
supersymmetric Yang-Mills theory for the 
gauge group $SU(2)$\cite{sw}. We will now summarize the main 
aspects of this solution, stressing the features that are 
relevant to Donaldson theory. For a review of the Seiberg-Witten solution, 
one can see reference\cite{luis}. 

1) Classically, the gauge symmetry is broken by the VEV of the gauge-invariant 
field $u=\hbox{\rm Tr}(B^2)$, and is restored at $u=0$, where the $W^{\pm}$ 
bosons are massless. The classical moduli space of vacua is parametrized by 
the complex variable $u$ and is singular at the origin.

2) Quantum-mechanically, the $SU(2)$ symmetry is never restored and the 
theory stays in the Coulomb phase throughout the complex $u$-plane. The 
low-energy description is then given by an $N=2$ abelian vector 
multiplet ${\cal A}$ (whose scalar component will be denoted by $a$) 
and an $N=2$ prepotential ${\cal F}({\cal A})$. 

3) There is a discrete anomaly-free symmetry ${\bf Z}_8 \subset U(1)_{\cal R}$
 which acts as a ${\bf Z}_2$ symmetry on the $u$-plane (as $u$ has 
${\cal R}$-charge 4). The quantum moduli space of vacua has two singularities 
at $\pm \Lambda^2$ (where $\Lambda$ is the dynamically generated scale of 
the theory), related by the ${\bf Z}_2$ symmetry. These singularities are 
associated to extra degrees of freedom becoming massless. At $u=\Lambda^2$ 
there is a massless monopole with quantum numbers $(n_m, n_e)=(1,0)$, and 
at $u=-\Lambda^2$ there is a massless dyon with numbers $(1,-1)$. 

4) The low-energy effective theory has an $SL(2, {\bf Z})$ duality which 
relates different descriptions. Massless states at the singularities are 
described by $N=2$ matter hypermultiplets, and in each case one has to use the 
appropriate description of the abelian field using $SL(2,{\bf Z})$ 
transformations. For the monopole singularity one takes the magnetic, $S$-dual
 photon ${\cal A}_D$, while for the dyon one takes the dyonic photon 
${\cal A}_D-{\cal A}$. The scalar component of ${\cal A}_D$, $a_D$, gives 
the monopole mass through the BPS equation $M={\sqrt 2}|a_D|$. Therefore 
$a_D(\Lambda^2)=0$. In the same way, $(a_D-a)(-\Lambda^2)=0$ 

Now that we have a precise description of the IR physics of $N=2$ supersymmetric 
Yang-Mills theory, we can try to compute the Donaldson invariants 
in this regime. Notice that at every point of the quantum moduli space of vacua 
we have a different theory. The twisting of these theories gives precisely 
the effective TQFTs associated to Donaldson 
theory. In general, we will have two different kinds of theories.

 At a generic point in the $u$-plane, the resulting theory is just the 
abelian version of Donaldson theory, as the effective description 
 there includes 
only an abelian $N=2$ vector multiplet. In the same way, these theories will 
simply describe the moduli space associated to abelian instantons on $X$. 

 Near the singularities, an $N=2$ matter hypermutiplet, coupled to the 
abelian $N=2$ vector multiplet, must be also considered. This hypermultiplet 
describes the additional light degree of freedom (monopole or dyon) near 
$u=\pm \Lambda^2$. To describe then the twisted theory at the 
singularities, we have to perform 
the twist of the $N=2$ matter hypermultiplet. The hypermultiplet contains 
a complex scalar isodoublet $q=(q_1, q_2)$, fermions $\psi_{q \alpha}$, 
$\psi_{{\tilde q} \alpha}$, ${\overline \psi}_{q \dot \alpha}$, 
${\overline \psi}_{{\tilde q} \dot \alpha}$, and a complex scalar isodoublet 
auxiliary field $F_i$. The fields $q_i$, $\psi_{q \alpha}$, 
${\overline \psi}_{{\tilde q} \dot \alpha}$ and $F_i$ are in the fundamental 
representation of the gauge group, while the fields $q^{i \dagger}$, 
$\psi_{{\tilde q} \alpha}$, ${\overline \psi}_{q \dot \alpha}$ and 
$F^{i \dagger}$ are in the conjugate representation. As the fields $q_i$ are
charged under the $SU(2)_{I}$ symmetry, after  the twist they become
positive-chirality spinors $M_\alpha$. These fields  are in fact
sections of the bundle $S^{+} \otimes L$, where  $L$ is the line bundle
associated to the $U(1)$ gauge field, and the  resulting twisted theory will not
be well defined on a general four-manifold.  However, it has been shown in
\cite{sdual} that the {\it dual} line bundle  for the magnetic photon is not
globally defined as a line bundle, but rather  defines a ${\rm
Spin}^c$-structure. This means, on one hand, that $L^2$ is well  defined as a
line bundle, and on the other hand that the tensor product  $S^{+} \otimes L$ is
also well defined. This tensor product  bundle is called the 
(positive-chirality) complex  spinor bundle associated to the ${\rm
Spin}^c$-structure defined by $L$.  Once this  subtlety has been understood, we
can already read from the twisted action  the equation for the section $s$. In
this case, the configuration space for a  given choice of the ${\rm
Spin}^c$-structure is 
\be
{\cal X}={\cal A}_{L^2} \times \Gamma(X, S^{+} \otimes L),
\label{confimon}
\ee
where the first factor refers to the $U(1)$ connections on the line bundle 
$L^2$, and the second factor denotes the sections of the positive-chirality 
complex spinor bundle. The vector space ${\cal F}$ is now,
\be 
{\cal F}=\Omega^{2,+}(X) \oplus \Gamma(X, S^{-} \otimes L),
\label{fibramon}
\ee
where the second factor denotes the sections of the negative-chirality 
complex spinor bundle. The section $s$ has then two components, and its 
zero locus is defined by the equations
\bea
{1 \over 2} F^{+}_{\alpha \beta} + i {\overline M}_{(\alpha} M_{\beta)} &=&0, 
\nonumber\\ 
D^{\dot \alpha \alpha}_{L^2}M_{\alpha} &=& 0.
\label{swmon}
\eea
 These are the celebrated {\it Seiberg-Witten monopole equations}, and 
were introduced in reference\cite{abm}. We have written them using spinor
notation.  $F^{+}_{\alpha \beta}$ denotes 
the self-dual part of the curvature associated to the connection $A$ on $L^2$. 
The $1/2$ factor arises because what appears in the action is the curvature 
of the non-globally defined bundle $L$. We also denoted 
${\overline M}_{(\alpha} M_{\beta)}={\overline M}_{\alpha} M_{\beta}
+{\overline M}_{\beta} M_{\alpha}$. In the second equation, $D_{L^2}$ denotes 
the Dirac operator associated to the positive-chirality complex spinor bundle. 
The term ${\overline M}M$ in the first equation comes from the $D$-terms 
of $N=2$ QED. The solutions to the equations (\ref{swmon}), 
modulo gauge transformations, 
form a compact moduli space of dimension 
\be
\hbox{\rm dim} \,\ {\cal M}_{\rm SW}={1 \over 4} (x^2 -2 \chi -3 \sigma),
\label{dimab}
\ee
where $x=-c_1(L^2)$. From now on we will denote ${\rm Spin}^c$-structures by 
this $x$ (recall that line bundles can 
be completely classified by their first Chern class). 
One can define invariants just like in 
Donaldson theory, and when the dimension of the moduli space of abelian 
monopoles is zero, one can compute the partition function of the corresponding 
TQFT. The resulting invariant, associated to a 
${\rm Spin}^c$-structure $x$ on $X$, is called the Seiberg-Witten 
invariant, and 
will be denoted by $n_x$. The 
${\rm Spin}^c$-structures $x$ for which the corresponding 
Seiberg-Witten invariants is 
non-zero are called {\it basic classes}. They verify in particular 
$x^2=2 \chi +3 \sigma$. The TQFT associated to the moduli problem in (\ref{swmon}) was 
described in detail in reference\cite{abmono}. The mathematical aspects of the 
Seiberg-Witten monopole equations have been reviewed in reference\cite{matilde}

Now we can try to evaluate the generating functional (\ref{gendon}). If we 
do the path integral in terms of the effective description, one has to 
integrate over the $u$-plane and then perform the path integral for each 
of the different effective theories. Let us discuss carefully the steps in this 
computation:

1) First of all, if we assume (as we did before) that $b_2^+>1$, an important 
simplification occurs. Recall that, if $u \not= \pm \Lambda^2$, then 
the effective TQFT describes the moduli space 
of abelian instantons on $X$. But if $b_2^+>1$ there are no abelian 
solutions to the ASD equations, and this moduli spaces is empty. Therefore, the 
only contributions to the path integral come from the singularities.

2) Let's focus now on the monopole singularity at $u=\Lambda^2$. 
We know that the TQFT there describes the Seiberg-Witten monopoles, {\it i.e.}, the 
solutions to (\ref{swmon}). But we must also compute VEVs of operators, 
and then we must write the observables of Donaldson-Witten theory in terms 
of operators of the effective, abelian theory. To obtain the structure of this 
expansion we will use the expansion in the untwisted, physical theory, together 
with the descent equations in the topological abelian theory\cite{wiper}. The 
descent equations for the abelian monopole theory can be found in 
\cite{abmono}. Near the monopole singularity, the $u$ variable has 
the expansion
\be
u(a_D)= \Lambda^2 + \Big( {du \over da_D} \Big)_{0}a_D + {\rm 
higher \,\  order}, 
\label{expan}
\ee
where $(du / da_D)_{0}= -2i \Lambda$\cite{sw}, 
and ``higher order" means operators of 
higher dimensions in the expansion. The field $a_D$ corresponds to the field 
$\phi_D$ of the topological abelian theory\cite{abmono}, while the 
gauge-invariant parameter $u$ corresponds to the observable (\ref{four}) 
(after complex conjugation). In 
terms of observables of the corresponding twisted theories, the expansion 
(\ref{expan}) reads
\be
{\cal O}= \langle {\cal O} \rangle -{1 \over \pi} \langle V \rangle \phi_D 
+ {\rm higher \,\ order}, 
\label{descef}
\ee
where $\langle {\cal O} \rangle$, $\langle V \rangle$ are real $c$-numbers 
which should be related to the values of $u(0)$, $(du / da_D)_{0}$ in the 
untwisted theory. From the observable ${\cal O}$ one can obtain the observable 
${\cal O}^{(2)}$ by the descent procedure, and applying this procedure in the 
abelian TQFT to the r.h.s. of (\ref{descef}), we obtain:
\be
{\cal O}^{(2)}=-{1 \over \pi}\langle V \rangle F + {\rm higher \,\ order},
\label{descedos}
\ee
where $F$ is the dual electromagnetic field (associated to the magnetic 
monopole). In particular, taking into account that $x=-c_1(L^2)=-[F]/\pi$, 
where $[F]$ denotes the cohomology class of the two-form $F$, we finally obtain:
\be
I(\Sigma)=\langle V \rangle (\Sigma \cdot x) + {\rm higher \,\ order},
\label{final}
\ee
where the dot denotes the pairing between $2$-cohomology and $2$-homology. 
From the point of view of TQFT, higher 
dimensional terms should not contribute in the expansion, because of the 
invariance of the theory under rescalings of the metric. Manifolds for which 
this is indeed true are called of {\it simple type}. It seems that all simply 
connected four-manifolds with $b_2^+>1$ are of simple type, according to our 
expectations. 

We can now evaluate the correlation function in (\ref{gendon}) in terms of a 
path integral in the low energy twisted theory, for the monopole singularity.
 First of all, we have to sum 
over the topological sectors in the abelian twisted theory, {\it i.e.}, 
over all the ${\rm Spin}^c$-structures $x$. For manifolds of simple type, 
the expansion of the observables only includes $c$-numbers, as we can see 
in (\ref{descef}) and (\ref{final}), and, possibly, 
contact terms for the operators $I(\Sigma)$ (see reference\cite{wijmp} for a
discussion  on this point). These terms give the intersection form of the
manifold.  Denoting $v= \sum_a \alpha_a I(\Sigma_a)$, the intersection form
appears as  $\gamma v^2 = \gamma \sum_{a,b}\alpha_a \alpha_b (\Sigma_a,
\Sigma_b)$, where $\gamma$  is a real number. As the operators are now
$c$-numbers, the computation  of the path integral just gives the partition
function of the theory for  each ${\rm Spin}^c$-structure $x$. In particular, 
only the basic classes appear in the sum over topological sectors. 
We then obtain, for the first singularity,
\be
C \hbox{\rm exp}(\gamma v^2 +\mu \langle {\cal O} \rangle) \sum_{x} n_{x} {\rm e}^ 
{\langle V \rangle v \cdot x},
\label{singu}
\ee
The constant $C$ appears in the comparison of the macroscopic and microscopic 
path integrals, after fixing the respective normalizations. Some steps to 
determine it have been given in reference\cite{sdual}. 

3) We should now obtain the contribution from the dyon singularity at 
$u=-\Lambda^2$. But to do this we only need to implement the ${\bf Z}_8$ 
symmetry of the underlying $N=2$ supersymmetric Yang-Mills theory, as this 
symmetry in the $u$-plane relates the two vacua (see reference\cite{wijmp}). With 
our choice of twisting, and the correspondence between physical and twisted 
fields, we see that the ${\cal O}$ operator has ${\cal R}$-charge $4$, 
while the $I(\Sigma)$ operator 
has ${\cal R}$-charge $2$, and they transform under the generator $\alpha$
of the ${\bf Z}_8$ symmetry as 
\be
\alpha{\cal O} \alpha^{-1}= -{\cal O}, \quad \alpha I(\Sigma) \alpha^{-1} 
=i I(\Sigma). 
\label{zetados}
\ee
Recall that ${\bf Z}_8$ is the anomaly-free discrete subgroup of the 
$U(1)_{\cal R}$ symmetry. But on a curved manifold there is a gravitational 
contribution to the anomaly that can be computed in a standard way 
using the index theorem. The path integral 
measure around the singularity at $u=-\Lambda^2$ picks then a phase of the form 
\be
\hbox{\rm exp}\Big({i \pi \over 4}\hbox{\rm dim}\,\  {\cal M}_{ASD}\Big)=
i^{\Delta}, \label{fase}
\ee 
where ${\Delta}=(\chi + \sigma)/4$, and we have used the fact that 
$\hbox{\rm dim}\,\  {\cal M}_{ASD}$ is an even integer. 

Taking all this information into account, 
the correlation function of Donaldson theory is given by the sum of the 
monopole contribution, given in (\ref{singu}), and its ${\bf Z}_8$ transform, 
which gives the dyon contribution. The final result is  
\be
C \Big( \hbox{\rm exp}(\gamma v^2 +\mu \langle {\cal O} \rangle) 
\sum_{x} n_{x} {\rm e}^ { \langle V \rangle v\cdot x}+
i^{\Delta}\hbox{\rm exp}(-\gamma v^2 - \mu \langle {\cal O} \rangle) 
\sum_{x} n_{x} {\rm e}^ {-i  \langle V \rangle v \cdot x} \Big),
\label{vevdef}
\ee
and the unknown universal constants can be obtained after comparing to explicit 
results in the mathematical literature,
\be
C=2^{1+ (7 \chi + 11 \sigma)/4}, \quad \gamma=1, \quad {\cal O}=2, \quad 
\langle V \rangle =1.
\label{const}
\ee
This remarkable result for the Donaldson invariants in terms of Seiberg-Witten 
invariants was first obtained in reference\cite{abm}, although in the case of 
K\"ahler manifolds it was previously obtained in reference\cite{wijmp}, using 
exact results in $N=1$ supersymmetry. The fact that 
Donaldson invariants have the structure shown in (\ref{vevdef}) for 
some $2$-dimensional cohomology classes $x$ is precisely the content of the 
Kronheimer-Mrowka structure theorem\cite{km}.

\section{Non-Abelian Monopoles}

So far we have discussed two different moduli problems in four-dimensional 
topology, the ASD instanton equations and the Seiberg-Witten monopole 
equations. There is a natural generalization of these moduli problems 
which is non-abelian and includes spinor fields. These are the {\it non-abelian 
monopole equations}, introduced in reference\cite{nabm} from the point of view of
the  Mathai-Quillen formalism and as a generalization of Donaldson theory. They 
have been also considered in reference\cite{park} as well as in the mathematical 
literature\cite{oko,tele,pt,oscar}. 

To introduce these equations in the case of $G=SU(N)$ and the 
monopoles in the fundamental representation ${\bf N}$, consider a Riemannian 
four-manifold $X$ together 
with an $SU(N)$-bundle $P$ and an associated vector bundle $E$ in the 
fundamental representation. Suppose for simplicity that the manifold is 
${\rm Spin}$, and consider a section $M_{\alpha}^i$ of $S^{+}\otimes E$. The 
non-abelian monopole equations read in this case,
\bea
F^{+ij}_{\alpha \beta} + i ( {\overline M}^j_{(\alpha} M^i_{\beta)}-
{\delta^{ij} \over N}{\overline M}^k_{(\alpha} M^k_{\beta)})
 &=&0, 
\nonumber\\ 
(D^{\dot \alpha \alpha}_{E}M_{\alpha})^i &=& 0.
\label{namon}
\eea
These equations can be analyzed both from the mathematical and the 
physical point of view\cite{nabm,last,tesis,baryon,eq}. We will here 
summarize some of the most salient features for the $SU(2)$ case:

1) The moduli space of non-abelian 
monopoles associated to (\ref{namon}) includes the moduli space of ASD instantons 
as a subset, and in fact the usual conditions to have a well-defined moduli 
problem (like the reducibility) are essentially the same as in Donaldson 
theory. The description of this moduli space can be made more precise in the 
case of K\"ahler manifolds\cite{nabm,oko}.

2) The equations \ref{namon} arise in the twisting of $N=2$ QCD with one 
massless hypermultiplet ($N_f=1$)\cite{last,park,tesis}. They 
can be also defined on general four-manifolds using 
${\rm Spin}^c$-structures, as in the abelian case\cite{park,oko,oscar,pt}.
 It has 
been recently shown that, from the point of view of the twisted $N=2$ QCD 
theory, the inclusion of ${\rm Spin}^c$-structures corresponds to an extended 
twisting procedure associated to the gauging of the baryon number\cite{baryon}. 

3) The moduli space of non-abelian monopoles has a natural $U(1)$ action which 
acts as a rotation on the monopole fields. The fixed points of this action 
are essentially the moduli space of ASD instantons and the moduli space 
of abelian Seiberg-Witten monopoles. This has opened the way to a mathematical 
proof of the equivalence of both theories using localization techniques 
\cite{pt,oko}, and some promising and concrete results in this direction 
have been recently obtained\cite{okodos}. From the point of view 
of the Mathai-Quillen formalism, the $U(1)$ action makes the bundle ${\cal E}$ 
in (\ref{fibrao}) a $U(1)$-equivariant bundle and one can obtain a general
expression for  the equivariant extension of the Thom form in this
formalism\cite{eq}. For the  non-abelian monopole theory, the TQFT associated 
to this extension is precisely twisted $N=2$ QCD with one massive flavour. 
It is tempting to think that the physics of the massive theory could shed 
some light on the localization problem.

4) Because of the equivalence of the non-abelian monopole theory and $N=2$ 
supersymmetric QCD, one can compute the topological invariants associated 
to their moduli 
space using the techniques of the previous section and the Seiberg-Witten 
solution for the $N_f=1$ theory\cite{last}. They can also 
be computed in the K\"ahler case using exact results for 
superpotentials in $N=1$ supersymmetry and some additional assumptions. In the 
$N_f=1$ case, the quantum moduli space of vacua has three singularities related 
by a ${\bf Z}_3$ symmetry, and we have then three different contributions. The 
observables have the same structure as in the Donaldson-Witten theory 
(although they have a different geometrical content, as they implicitly 
include contributions from the matter fields). The effective description 
of the theory is again in terms of a dual $N=2$ QED with one massless 
hypermultiplet, and the main result of this computation 
is that the invariants associated to the non-abelian monopole theory can 
be expressed in terms of Seiberg-Witten invariants. The generating 
function for manifolds with $b_2^+ >1$ is\cite{last}:
\begin{eqnarray}
& &\langle {\rm exp} (\sum_{a}\alpha_{a}I(\Sigma_{a})+\mu {\cal O})
\rangle \nonumber \\
&=&C \Bigg( {\rm exp} (\gamma v^2 
+\mu \langle {\cal O} \rangle) \sum_{x} n_x {\rm
exp}(\langle V \rangle v \cdot x)  \nonumber \\
&+&{\rm e}^{-{\pi i \over 6}\sigma}
{\rm exp}\Big( -{\rm e}^{-{\pi i \over 3}}
(\gamma v^2 +\mu \langle{\cal O} \rangle)
\Big) \sum_{x} n_x {\rm exp}({\rm e}^{-2\pi i \over 3} \langle V
\rangle v \cdot x) \nonumber \\
&+&{\rm e}^{-{\pi i \over 3}\sigma}
{\rm exp}\Big( -{\rm e}^{{\pi i \over 3}}
(\gamma v^2 +\mu \langle{\cal O} \rangle)
\Big) \sum_{x} n_x {\rm exp}({\rm e}^{-4\pi i \over 3} \langle V
\rangle v \cdot x) \Bigg). \nonumber \\
\label{fernanda}
\end{eqnarray}
where unknown constants appear as in the pure Donaldson-Witten case.

The result (\ref{fernanda}) suggests that moduli problems in four-dimensional
topology can be  classified 
in {\it universality classes} associated to the effective low-energy description 
of the underlying physical theory. One important question that should be
addressed is how large is the set of moduli spaces which admit a description in
terms of Seiberg-Witten invariants. It is very likely that in the search for
this set new types of invariants will be found leading 
to new universality classes.

We will end comparing the situation found in four dimensions to the one in
three. In the latter, it turns out that all the Chern-Simons invariants can be
expressed in terms of Vassiliev invariants. These invariants play the role of
the set of universality classes for this case. As in the four-dimensional case,
the analysis of the TQFT from a perturbative and a non-perturbative point of
view  establishes their relation. In the four-dimensional case
Seiberg-Witten invariants constitute the first universality class of a
presumably large set of invariants. At present, the construction of this set is
one of the most important challenges in TQFT.

\vspace{4 mm}

\begin{center}
{\bf Acknowledgments}
\end{center}

\vspace{4 mm}

One of us (J.M.F.L) would like to thank the organizers
of the ``Workshop on Frontiers of Field Theory,
Quantum Gravity and String Theory" for their kind invitation. This work was
supported in part by DGICYT under grant PB93-0344.

\vspace{4 mm}

\begin{center}
{\bf References}
\end{center}

%\vspace{2 mm}


\begin{thebibliography}{99}


\def\np{Nucl. Phys.}
\def\pl{Phys. Lett.} 
\def\pre{Phys. Rep.} 
\def\prl{Phys. Rev. Lett.}
\def\pr{Phys. Rev.} 
\def\ap{Ann. Phys.} 
\def\cmp{Comm. Math. Phys.}
\def\ijmp{Int. J. Mod. Phys.} 
\def\mpl{Mod. Phys. Lett.} \def\lmp{Lett. Math. Phys.} 
\def\bams{Bull. AMS} \def\am{Ann. of Math.} 
\def\jpsc{J. Phys. Soc. Jap.} \def\topo{Topology} 
\def\ijm{Int. J. Math.}
\def\knot{Journal of Knot Theory and Its Ramifications} 
\def\jmp{J. Math. Phys.} 
\def\jgp{J. Geom. Phys.} 
\def\jdg{J. Diff. Geom.}
\def\plms{Proc. London Math. Soc.}
\def\mrl{Math. Res. Lett.}
\def\inma{Invent. Math.}
\def\tam{Trans. Am. Math. Soc.}


\bibitem{top} M. Alvarez 
and J.M.F. Labastida, {\sl\pl} {\bf B315} (1993) 251;
{\sl\np} {\bf B437} (1995) 356.

\bibitem{alla} M. Alvarez and J.M.F.
Labastida, {\sl\np} {\bf B395} (1993) 198; {\sl\np} {\bf B433} (1995) 555; 
Erratum, {\sl\np} {\bf B441} (1995) 403.

\bibitem{esther}
M. Alvarez,  J.M.F. Labastida and E. Perez, {\sl \np} {\bf B488} (1997) 677.

\bibitem{luis} L. \'Alvarez-Gaum\'e and S.F. Hassan, hep-th/9701169; 
W. Lerche, hep-th/9611190; A. Bilal, hep-th/9601007.

\bibitem{ans} D. Anselmi and P. Fr\`e, {\sl\np} {\bf B392} (1993) 401;
{\sl\np} {\bf B404} (1993) 288; {\sl\np}
{\bf B416} (1994) 25; {\sl\pl} {\bf B347} (1995) 247.

\bibitem{ahs}
M.F. Atiyah, N.H. Hitchin and I.M. Singer, {\it Proc. R. Soc. Lond} {\bf A362} 
(1978) 425. 

\bibitem{jeffrey} M. Atiyah and L. Jeffrey, {\sl J. Geom. Phys.} {\bf 7} (1990)
119.

\bibitem{natan} D. Bar-Natan, ``Perturbative aspects of Chern-Simons
topological quantum field theory", Ph. D. Thesis, Princeton University,
1991.

\bibitem{barnatan} D. Bar-Natan, {\sl\topo} {\bf 34} (1995) 423.

\bibitem{bilin} J.S. Birman and X.S. Lin, {\sl \inma} {\bf 111} (1993) 225;
J.S. Birman, {\sl\bams} {\bf 28} (1993) 253. 

\bibitem{thompson} D. Birmingham, M. Blau, M. Rakowski and G. Thompson, {\sl
Phys. Rep.} {\bf 209} (1991) 129.

\bibitem{oscar}
S. Bradlow and O. Garc\'\i a-Prada, ``Non-abelian monopoles and vortices", 
alg-geom/9602010.

\bibitem{moore} S. Cordes, G. Moore and S. Rangoolam, 
``Lectures on  2D Yang-Mills Theory, Equivariant Cohomology and
Topological Field Theory", in {\it Fluctuating geometries in statistical 
mechanics and field theory}, Les Houches Session LXII, F. David, P. Ginsparg 
and J. Zinn-Justin, eds. (Elsevier, 1996) p. 505.

\bibitem{donald}S. K. Donaldson, {\sl\topo} {\bf 29} (1990) 257.

\bibitem{gmm} E. Guadagnini, M.
Martellini and M. Mintchev, {\sl\pl} {\bf B227} (1989) 111;
{\sl\pl} {\bf B228} (1989) 489; {\sl\np} {\bf B330} (1990) 575.

\bibitem{park} S. Hyun, J. Park and J.S. Park, 
{\sl\np} {\bf B453} (1995) 199.

\bibitem{jones} V.F.R. Jones, {\sl \bams} {\bf 12} (1985) 103;
{\sl\am} {\bf 126} (1987) 335.

\bibitem{rocek} A. Karlhede and M. Ro\v cek, {\sl\pl} {\bf B212} (1988) 51.

\bibitem{kont} M. Kontsevich, {\sl Advances in Soviet Math.} {\bf 16}, Part 2
(1993) 137.

\bibitem{km} P.B. Kronheimer and T.S. Mrowka, {\sl \bams} {\bf 30} (1994) 215.


\bibitem{abmono} J.M.F. Labastida and M. Mari\~no,
{\sl\pl} {\bf B351} (1995) 146.

\bibitem{nabm} J.M.F. Labastida and M. Mari\~no, {\sl\np} {\bf B448} (1995)
373.  

\bibitem{last} J.M.F. Labastida and M. Mari\~no, 
{\sl\np} {\bf B456} (1995) 633.

\bibitem{eq} J.M.F. Labastida and M. Mari\~no, ``Twisted $N=2$ supersymmetry 
with central charge and equivariant cohomology", hep-th/9603169, to appear
in {\it Comm. Math. Phys.}

\bibitem{baryon} J.M.F. Labastida and M. Mari\~no, ``Twisted baryon number 
in $N=2$ supersymmetric QCD", hep-th/9702054, to appear in {\sl\pl}

\bibitem{matilde} M. Marcolli, ``Notes on Seiberg-Witten gauge theory",
dg-ga/9509005; J.W. Morgan, {\it The Seiberg-Witten equations 
and applications to the topology of smooth four-manifolds}, Princeton 
University Press, 1996.

\bibitem{tesis}
M. Mari\~no, The geometry of supersymmetric gauge theories in four 
dimensions, hep-th/9701128.  
 
\bibitem{mathai}
V. Mathai and D. Quillen, {\sl Topology} {\bf 25} (1986) 85.  

\bibitem{oko}
C. Okonek and A. Teleman, {\it Int. J. Math.} {\bf 6} (1995) 893; 
{\sl\cmp} {\bf 180} (1996) 363.

\bibitem{okodos}
C. Okonek and A. Teleman, ``Recent developments in Seiberg-Witten theory 
and complex geometry", alg-geom/9612015.   

\bibitem{pt}
V. Pidstrigach and A. Tyurin, ``Localisation of the Donaldson invariants 
along Seiberg-Witten classes", dg-ga/9507004.

\bibitem{sw} N. Seiberg and E. Witten, {\sl\np} {\bf B426} (1994) 19;
Erratum, {\sl\np} {\bf B430} (1994) 485; {\sl\np} {\bf B431} (1994) 484.

\bibitem{tele}
A. Teleman, ``Non-abelian Seiberg-Witten theory and projectively 
stable pairs", alg-geom/9609020. 

\bibitem{vass} V. A. Vassiliev, ``Cohomology of knot spaces", {\it Theory of
singularities and its applications}, {\sl Advances in Soviet Mathematics},
vol. 1, {\sl Americam Math. Soc.}, Providence, RI, 1990, 23-69.

\bibitem{tqft} E. Witten, {\sl\cmp} {\bf 117} (1988) 353.  

\bibitem{tsm} E. Witten, {\sl\cmp} {\bf 118} (1988) 411. 

\bibitem{csgt} E. Witten,  {\sl\cmp} {\bf 121} (1989) 351.

\bibitem{wijmp} E. Witten, {\sl \jmp} {\bf 35} (1994) 5101.


\bibitem{abm} E. Witten, {\sl\mrl} {\bf 1} (1994) 769.

\bibitem{wiper}
E. Witten, private communication.

\bibitem{sdual}
E. Witten, ``On $S$-duality in abelian gauge theory", hep-th/9505186.

\end{thebibliography}
\end{document}